%% file: text.tex
\documentclass[10pt,conference]{IEEEtran}
\IEEEoverridecommandlockouts
\usepackage{cite}
\usepackage{amsmath,amssymb,amsfonts}
\usepackage{algorithmic}
\usepackage{graphicx}
\usepackage{textcomp}
\usepackage{xcolor}

\usepackage{booktabs}
\usepackage[most]{tcolorbox}

\def\BibTeX{{\rm B\kern-.05em{\sc i\kern-.025em b}\kern-.08em
    T\kern-.1667em\lower.7ex\hbox{E}\kern-.125emX}}
\begin{document}

\title{Understanding the context of IoT software systems in DevOps}

\author{\IEEEauthorblockN{Igor Muzetti Pereira}
\IEEEauthorblockA{\textit{Department of Computing and Systems} \\
\textit{University Federal of Ouro Preto}\\
João Monlevade, Brazil\\
igormuzetti@ufop.edu.br}
\and
\IEEEauthorblockN{Tiago Garcia de Senna Carneiro}
\IEEEauthorblockA{\textit{Departament of Computing} \\
\textit{University Federal of Ouro Preto}\\
Ouro Preto, Brazil \\
tiago@ufop.edu.br}
\and
\IEEEauthorblockN{Eduardo Figueiredo}
\IEEEauthorblockA{\textit{Department of Computer Science} \\
\textit{University Federal of Minas Gerais}\\
Belo Horizonte, Brazil \\
figueiredo@dcc.ufmg.br}
}

\maketitle

\begin{abstract}
The growing demand for connected devices and the increase in investments in the Internet of Things (IoT) sector induce the growth of the market for this technology. IoT permeates all areas of life of an individual, from smartwatches to entire home assistants and solutions in different areas. The IoT concept is gradually increasing all over the globe. IoT projects induce an articulation of studies in software engineering to prepare the development and operation of software systems materialized in physical objects and structures interconnected with embedded software and hosted in clouds. IoT projects have boundaries between development and operation stages. This study search for evidence in scientific literature to support these boundaries through Development and Operations (DevOps) principles. We rely on a Systematic Literature Review to investigate the relations of DevOps in IoT software systems. As a result, we identify concepts, characterize the benefits and challenges in the context of knowledge previously reported in primary studies in the literature. The main contributions of this paper are: (i) discussion of benefits and challenges for DevOps in IoT software systems, (ii) identification of tools, concepts, and programming languages used, and, (iii) perceived pipeline for this kind of software development.
\end{abstract}

\begin{IEEEkeywords}
DevOps, IoT Software Systems, Internet of Things Software Systems.
\end{IEEEkeywords}

\section{Introduction}  \label{sec:intro}
Development and operations teams have opposite goals and incentives. The goal of development teams is to respond to business needs quickly. On the other hand, the goal of the operational teams is to provide a stable, reliable, and secure service to the customer \cite{manualdevops}. To address these conflicts, organizations adopt a culture of collaboration between development and operations teams (DevOps) \cite{fenix}. DevOps allows companies to increase their workflow completed, increasing the frequency of deployments and the stability and robustness of the production environment \cite{renatasbes}. 

Internet of Things (IoT) solutions share these conflicts, as they require changes in known ways of construction and maintenance \cite{brunosbes}. IoT solutions are composed of addressable objects (things) that interact with each other and with users, detecting and processing data to achieve specific goals \cite{jacobson}. IoT solutions have different software components, such as clients and servers, with their frontend and backend. They also include the essential software to the hardware (e.g., firmware, drivers).

Internet of Things leads to an era in which, instead of just developing software, we need to design software systems that encompass multidisciplinarity and integrate different areas for the realization of successful products according to their purposes. This means that software is just one facet of IoT, which, along with others, is needed for IoT solutions \cite{motta}. Therefore, software engineering is adapting to the construction and maintenance of these complex IoT systems, seeking to reduce development costs and increase the quality of the final product \cite{sredevops}.

IoT software systems involve recent technologies that are questioning the traditional way of developing. Software engineering, as a discipline, has undergone constant changes since its conception. Several concepts, methods, tools, and patterns support software development \cite{sweebok}. It is then possible to consider the need to evolve existing development practices to support the construction of IoT software systems. 

Some secondary studies in the literature individually address the concepts of DevOps and IoT. From an analysis of related papers, although DevOps is perceived as an automation-centric approach, the cunning of human communication and collaboration needs a lot of attention to be used well \cite{devops1, devops2, devops3}. Regarding studies related to IoT, they identified and discussed several scenarios for applications, research challenges, and open questions to be faced for the construction and maintenance of IoT projects \cite{iot1, iot2}. We are not aware of similar secondary studies that investigate the use of DevOps for IoT systems.

Through a Systematic Literature Review (SLR), this study aims to identify, analyze, and interpret evidence available in the literature on the explicit relationship between DevOps and IoT software systems. We applied a qualitative analysis, including open coding technique, to answer three research questions. All results come from evidences in twenty-six studies found in this SLR.

As the main contributions, we identified DevOps concepts that are successfully applied in the development and operation of IoT software systems. These concepts are formed by sets of best practices and techniques used in projects of this nature. We also created two lists with the benefits and challenges addressed in the selected studies. We discussed a set of tools and programming languages perceived in the studies. Finally, we also propose a pipeline for DevOps in IoT formed by different stages. 

The remainder of this paper is organized as follows. Section 2 describes the SLR protocol we have followed. Section 3 presents the results of the conducted SLR. Section 4 deals with the practical implications of the findings of this study. Section 5  discusses threats to the study validity. Section 6 discusses related work. Section 7 concludes this study and points directions for future work.

\section{Systematic Literature Review} \label{sec:slr}
Systematic Literature Review (SLR) is a study that provides identification, analysis, and interpretation of evidence related to a specific topic \cite{wohlin}. The conduction of this study followed guidelines of Kitchenham \cite{kitchenham}. It has three phases: (i) Planning, (ii) Execution, and (iii) Analysis. The focus of this SLR is on the explicit relationship between DevOps and IoT software systems. 

\subsection{Planning} 
In this phase, we define: (i) the topic to be investigated, (ii) the electronic data sources used to search for primary studies, (iii) the search string to identify the relevant studies, (iv) the inclusion and exclusion criteria for selecting primary studies, and (v) timestamp in which this study has been conducted.

\textbf{Research Questions}. To identify the relationship between DevOps and IoT in software systems, we defined three Research Questions (RQs).

\

\textbf{RQ1.} What are the DevOps concepts applied to the context of IoT software systems? 

\

\textbf{RQ2.} What are the reported benefits of adopting DevOps in IoT software systems?

\

\textbf{RQ3.} What are the reported challenges of adopting DevOps in IoT software systems?

\

\textbf{Electronic Data Sources (EDS).} Different EDS can be used in literature reviews to search for primary studies. We used five of them: IEEE Xplore Digital Library, ACM Digital Library, ScienceDirect, Springer, and Wiley. These EDS belong to the category of publisher sites known as part of the group of important sources for easy retrieval of relevant and published literature on software engineering \cite{databases}. 

\textbf{Search String.} We conducted a pilot study with search strings composed of several different terms and applied them to each EDS. In the end, the consolidated search string was:\textit{ ((``internet of things'' OR iot) AND devops)}.

\textbf{Inclusion and Exclusion Criteria.} We applied the following inclusion criteria: English and complete papers published in the last five years. Exclusion criteria: duplicate and short papers (less than six pages) and book chapters. In this study, the strategy adopted was to select papers describing DevOps applied to IoT software systems after reading and analyzing the metadata (i.e., title, abstract, and keywords).

\textbf{Search Source.} We researched all studies published between 2015 and since the research process was carried out in the first half of 2020. We chose this period to focus on recently published work.

\subsection{Execution} 
This phase consists of (i) applying the search string in the selected EDS, (ii) identifying the primary studies, and (iii) selecting the papers found by using the inclusion and exclusion criteria. 


\textbf{Study Selection Process.} Three steps with a focus on the selection of relevant papers, discussed as follows: Step 1 - Inclusion and exclusion criteria. Criteria used to include the relevant studies and to discard those that are not relevant; Step 2 - Metadata reading. The strategy adopted to characterize the scope of this SLR; Step 3 - Snowballing. In our SLR, we run the backward snowballing \cite{snowballing}. 

\textbf{Data Extraction.} Research questions are the main factors from which information needs to be extracted. For each selected paper, we identified and extracted evidence about the benefits and challenges of DevOps applied to IoT software systems. We use the open coding technique and discussions between the authors to answer the research questions and to create categories \cite{opencoding2}. 

\textbf{Summarization.} It is worth mentioning that all selected studies were published between 2015 and half 2020, and the number of publications grew each year.

\section{Results} \label{sec:results}
This section aims to answer the RQs presented in Section 2.1.

\subsection{Overview of Primary Studies} \label{sec:overview}
The twenty-six selected studies were published in conferences, workshops, journals, and magazines, with a total of fourteen, five, five, and two, respectively. They were posted in twenty-two different venues. The greater number of studies published in conferences and workshops supports the novelty of these topics. The dispersion of these numbers may also indicate that there is still no specialized venue for the community to send this type of study (internet of things and software engineering).

To establish the set of research strategies used, we classified the selected studies according to the study categories of Wieringa et al. \cite{classificacao}. The classification was carried out through discussions between the authors. Validation research was the category with more studies (six). Solution proposals and experience papers have five studies. Evaluation and opinion papers came after with four studies each. Finally, we found two philosophical papers.

\subsection{RQ1. What are the DevOps concepts applied in the context of IoT software systems?} \label{sec:concepts}

Reading the papers that define the themes helped us to identify the concepts. These concepts guided the first iterations of open coding. Each paper was read, looking for evidence of each of the concepts. Tools, techniques, and reported practices were categorized according to their objectives. Regarding monitoring, DevOps measures use it to influence business decisions and find flaws in the product's life cycle.

\begin{figure}[!h]
\centering
\includegraphics[width=\columnwidth]{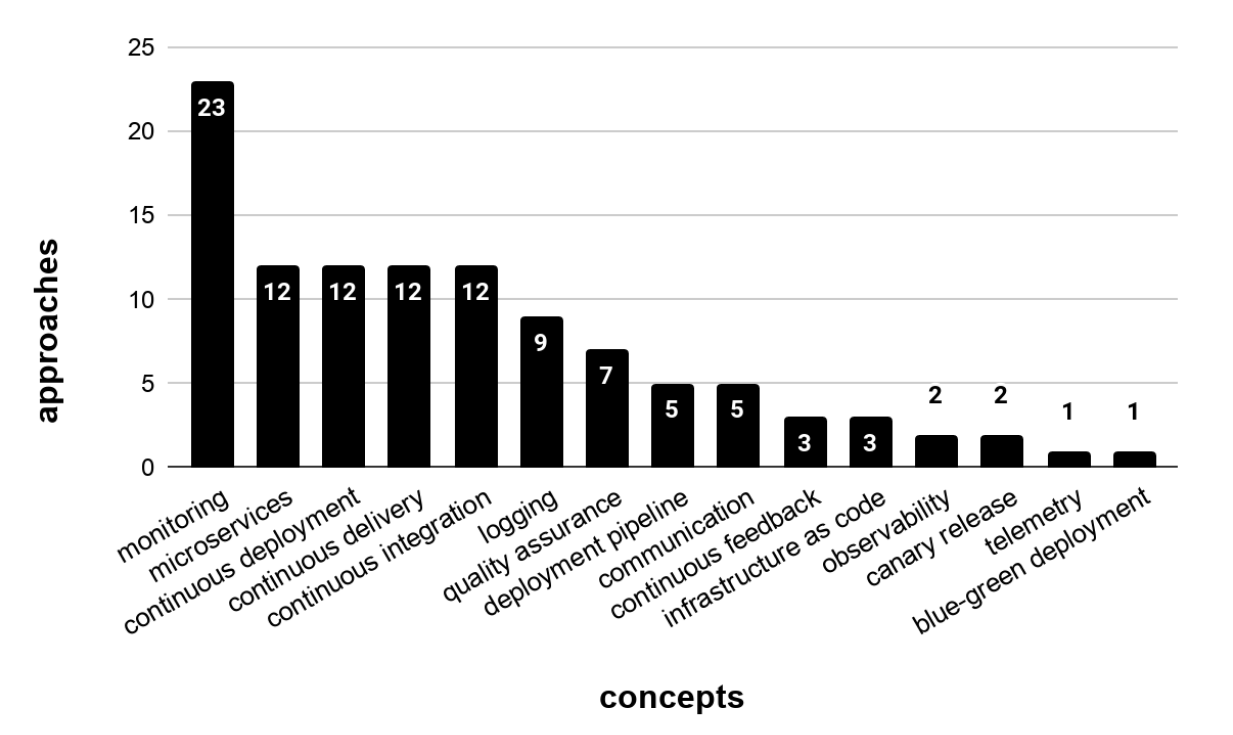}
\caption{DevOps concepts used in the studies.}~\label{fig:fig3}
\end{figure}

Figure~\ref{fig:fig3} presents the DevOps concepts used in IoT projects according to the selected studies. We observe a large number of concepts correlate with the automation of the deployment process. The automated deployment pipeline breaks up its build into stages. Each stage provides increasing confidence, usually at the cost of extra time. Early stages can find most problems yielding faster feedback, while later stages provide slower \cite{continousdelivery}.

The most addressed concept in the studies, present in twenty-three of them, was monitoring. In general, studies mention that, in an automated environment, it is essential to monitor the entire pipeline. However, it is clear that it is hard to monitor the entire deployment pipeline. Some tools used in the studies propose metrics to monitor development cycles, vulnerabilities in implementations, server performance, applications, and user activities. 

In addition to monitoring, other concepts were recurrently mentioned, such as microservices, deployment, delivery, and continuous integration covered in twelve studies. Regarding microservices, we consider that, in addition to contributing to faster deliveries, they support the creation of small teams (see Table~\ref{tab:tab5}). Deployment, delivery, and continuous integration shorten the feedback cycle, generating and incorporating value and quickly detecting failures.

Table~\ref{tab:tab4} presents forty-four tools distributed in the possible stages of DevOps according to their goals. It is essential to understand that some tools have their main features more related to a specific stage. 

\input{tables/Tools.tex}

However, they can have cross-functionalities to cover several DevOps stages. It is also important to note that many other tools can complement a DevOps deployment pipeline for IoT software systems. The deployment pipeline streamlines the process of delivering value. The identification of the stages is related to the common objective among the studies that apply DevOps in IoT projects: deliver software products in an automated way.

\begin{figure}[!h]
\centering
\includegraphics[width=\columnwidth]{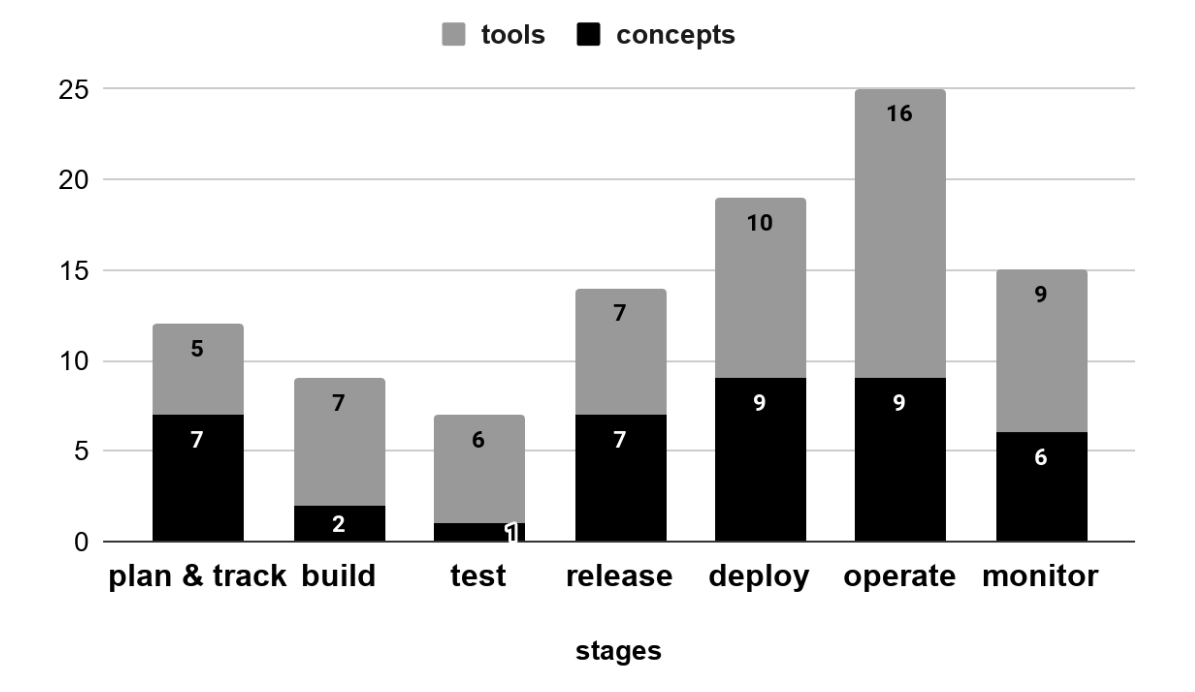}
\caption{Concepts and tools by DevOps stages.}~\label{fig:fig4}
\end{figure}

Figure~\ref{fig:fig4} presents a stacked column chart with the most discussed stages given the number of concepts and tools. The most reviewed stages were those that are mainly part of the operations cycle. This may have happened because the creation of DevOps is needed to provide greater agility in operations and infrastructure activities. These activities are essential in IoT software systems. On the other hand, little focus was given on the testing and build stages, suggesting a possible research gap \cite{infraopsagil}.

\begin{tcolorbox}[colframe=black!80, colback=white!15, boxrule=0.5pt,arc=0.2em,boxsep=-1mm]

\textbf{Answering RQ1 -} We identified fifteen different DevOps concepts and correlated them with the automation of the deployment process.
\end{tcolorbox}

\subsection{RQ2. What are the reported benefits of adopting DevOps in IoT software systems?}

\input{tables/Benefits.tex}

Table~\ref{tab:tab5} shows the benefits perceived with the use of DevOps in IoT software systems found in the selected studies. The second column contains the categories of benefits. The third column contains the references of studies that mention each benefit. The concepts in Section ~\ref{sec:concepts} are related to the benefits discussed in this section.

The benefit B01 is mentioned in eleven studies. With respect to B01, frameworks address different stages of a DevOps pipeline. For example, frameworks have been proposed to deploy applications in the clouds using DevOps; to evaluate microservice delivery strategies; to guarantee quality in IoT; reference architecture for multi-cloud IoT; a model-based framework for IoT; conceptual guidelines for the development and operation in IoT; orchestration model for serverless computing; DevOps model for edge resources and to control virtual infrastructure functions during DevOps in IoT.

In B02, the combination of Cloud and IoT technologies is known in the literature as a new IT paradigm, CloudIoT \cite{cloudiot}. The elasticity promoted by the cloud provides an illusion of unlimited resources in terms of storage and processing capacity. However, it cannot be considered as a source of infinite resources, as it depends on the demand for a complete set of customers. The use of clouds for back-end makes IoT takes advantage of this elasticity. Besides, because it is a more mature technology, the cloud has solved many problems that guarantee reliability, security, and privacy to CloudIoT solutions \cite{cloudcomputing}. The use of cloud computing in IoT projects that use DevOps for their construction provides high availability to services. It also provides greater agility to achieve requirements, such as scalability, elasticity, disaster recovery, and security. It also supports savings based on usage and transforms commercial capital expenditures into variable capital expenditures \footnote{https://aws.amazon.com/devops/what-is-devops/}. The application of blue-green deployment\footnote{https://martinfowler.com/bliki/BlueGreenDeployment.html} and canary-release\footnote{https://martinfowler.com/bliki/CanaryRelease.html} concepts can be achieved through cloud computing models.

In B03, packaging and isolating the application with the entire runtime environment makes it possible to migrate that container to other development, test or production environments. This also makes it possible for this container to communicate with any operating system of server without losing any aspect. B04 was conceived by discussions in papers that report that microservices promote an organization of small independent teams. The teams act within a defined and well-understood context and have the power to work independently and quickly, thus reducing delivery times. Microservices create modularity, individually, and can perform a single responsibility \cite{microservices}. They can be reused in different components of the solution and attributed to small teams. 

B05 refers to proposing mechanisms for the initial preparation steps to configure a new resource in the context of telecommunications. These mechanisms deal with the provisioning of devices that use cellular networks, such as allocating a new line, configuring the network components that allow completing calls, setting extra services such as SMS or e-mail, and, finally, associating all this with a chip. Platforms to support these networks are being increasingly designed \cite{ericsson}.

\begin{tcolorbox}[colframe=black!80, colback=white!15, boxrule=0.5pt,arc=0.2em,boxsep=-1mm]

\textbf{Answering RQ2 -} Five benefits were identified, explicitly discussed in different studies.
\end{tcolorbox}

\subsection{RQ3. What are the reported challenges of adopting DevOps in IoT software systems?}  

Table~\ref{tab:tab6} shows the challenges perceived with using DevOps on IoT. Some challenges were indicated as benefits achieved in other studies. Other challenges can also be considered open questions in the context of DevOps and IoT. The concepts in Section ~\ref{sec:concepts} are related to the challenges discussed in this section.

\input{tables/Challenges.tex}

C01 is often mentioned because there is a significant dependence on manual inspections and requirements for various validation processes, making instrumentation for collecting dynamic metrics difficult. C02 involves difficulties in configuring, updating firmware, and monitoring device status. About C03, we can see in the stacked column chart in Figure~\ref{fig:fig4} that there is an emphasis on studies of the operation cycle stages, which encourage the creation of tools and concepts to fill the gap of continuous feedback. 

The solution architecture is actively involved in the treatment of C04. The use of cloud computing is widely discussed in these studies to facilitate a broader scope of different non-functional requirements. The requirements mostly achieved through the clouds and discussed in the studies are scalability, elasticity, security, and performance \cite{cloudcomputing}. We agree with the conclusion of the study of Carvalho et al \cite{carvalhosbes}. The adequacy of several non-functional requirements presents several difficulties to be addressed. Negative correlations are one of those problems that developers must deal with during software development. These correlations mean that supporting one characteristic can negatively impact another. A known solution is through non-functional requirements catalogs (NFRs) that even cloud platforms usually provide.

The application of a deployment pipeline is taken into account in the studies as a challenge (C05). Often, the beginning of a DevOps implementation is the application of an automated CI/CD pipeline (Continuous Integration and Delivery). This change in the process brings about a cultural change for the team and is reported as a challenge. With respect to C06, this challenge is about developing continuous documentation. This can create a bottleneck and fail to offer higher value to customers. An approach in the literature, DevDocOps, proposes a method that can help to meet this challenge \cite{devdocops}.

C07 addresses the need for e-shaped professionals. This profile allows for quick workflows, as they are self-sufficient and do not create bottlenecks. IoT systems use different and recent technologies between hardware and software, which is why these professionals are wanted. The assessment of the quality of service offered is a challenge, according to C08. It is discussed that to increase the popularity of any service, Quality of Service (QoS) metrics must be defined so that the users can understand and express their needs. The heterogeneity and distribution between the components of an IoT solution make monitoring difficult.  C09 is mentioned in studies that explain the need for DevOps frameworks for IoT. In this study, the most discussed benefit (B01) deals with precisely this subject.

\begin{tcolorbox}[colframe=black!80, colback=white!15, boxrule=0.5pt,arc=0.2em,boxsep=-1mm]

\textbf{Answering RQ3 -} Nine challenges were identified and explicitly discussed in at least two different studies.
\end{tcolorbox}

\section{Discussion}
This section discusses the technical and practical implications identified in the studies. The analyzed papers present complementary views of the term DevOps. We understand that DevOps is the junction of the term Development + Operations. DevOps is a fluid, complex, and not so simple to define a concept. Many consider Patrick Debois's meeting with Andrew Shaffer in Toronto in 2008 to be ground zero. Still, a significant event is the presentation of John Allspaw and Paul Hammond, "10 Deploys a day" on Flickr \footnote{https://www.youtube.com/watch?v=ctMa7rFOHiU}, he influenced Debois to organize DevOpsDays. If we try to see the essence of all the lectures, papers, and books on the subject, three factors always permeate these different sources: people (collaboration), flow (pipeline, development cycle), and resilience (preparing systems to face failure situations).

The OWASP project \footnote{https://owasp.org/www-project-internet-of-things/} released a list of the top ten factors that impact the development of IoT. This list is intended to help manufacturers and developers better understand IoT's security issues and failures. A Gartner report \footnote{https://www.gartner.com/en/doc/3834669-implementing-and-executing-your-internet-of-things-strategy-a-gartner-trend-insight-report} concludes that IoT is the foundation on which digital businesses are being built. The report emphasizes three challenges that organizations must face at the beginning of their IoT transformation: building stakeholder confidence, choosing the right technology strategy, and executing that strategy. We believe that many of these factors and challenges can be addressed, including analyzing this study's contributions. It involves the adoption of DevOps practices in IoT software system projects.

\subsection{Most Studied Languages}

\begin{figure}[!h]
\centering
\includegraphics[width=\columnwidth]{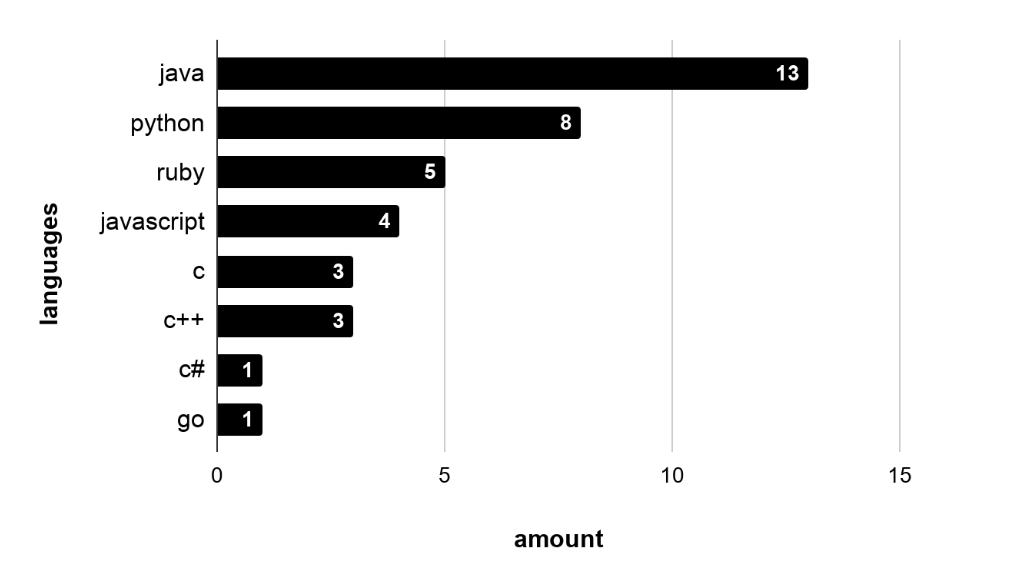}
\caption{Languages covered in the studies.}\label{fig:fig5}
\end{figure}

Figure~\ref{fig:fig5} presents a bar chart ordered according to the languages cited in the selected papers that are used in the studies. In total, eight different programming languages distributed in the selected studies were mentioned. The Java language was the most addressed with thirteen citations. Java has been around for a long time and is still widely used. The languages C and C ++ are used in IoT projects, but mainly they were discussed in studies that involved the construction of essential software for the hardware (firmware, driver).

To work with the operating tools, Python and Ruby stand out. Python is used extensively for backend code and scripts. Since DevOps is about automating multiple processes, Python can help implement CI/CD (continous integration and deployment) or configuration management efforts with Jenkins, Chef, Ansible, and other tools. Ruby is a scripting language that offers versatility. Although commonly used for Web development, it is also great for infrastructure management due to its flexibility. Both Python and Ruby have a large active community. They share many code examples, integrations, modules, and tutorials to disseminate knowledge \cite{devopslanguages}.

Although JavaScript is not as flexible as Python and Ruby, it is still common enough to add value in a DevOps environment. It was used in client and server codes. Some study designs are based on Node.js \cite{s24} \cite{s19} \cite{s11} \cite{s04} and \cite{s03}. Similar to Python and Ruby, JavaScript benefits from a vast user base. The C\# language is used in the software development toolkits for today's popular IoT development board \cite{s09}. The Go language little mentioned is the newest\footnote{https://golang.org/doc/}. It combines the speed of development in Python with the performance and security of languages like C or C ++. Go is a language compiled and focused on productivity and concurrent programming. Although its popularity as a whole, Go is increasing and is reasonably straightforward compared to others of the same type (e.g., C, C++). However, Go is not as simple as Python, which seems to be the main barrier. Perhaps, it can be used more in this context if it evolves to become more succinct. The amount of code that needs to be written for even the simplest things is enormous.

Eight programming languages were identified in the selected studies. They are present in different types of software that IoT software system solutions have. The decreasing order that the languages were addressed in the studies is Java, Python, Ruby, JavaScript, C, C++, C\#, and Go.

\subsection{Perceived Pipeline}
\begin{figure}[!h]
\centering
\includegraphics[width=\columnwidth]{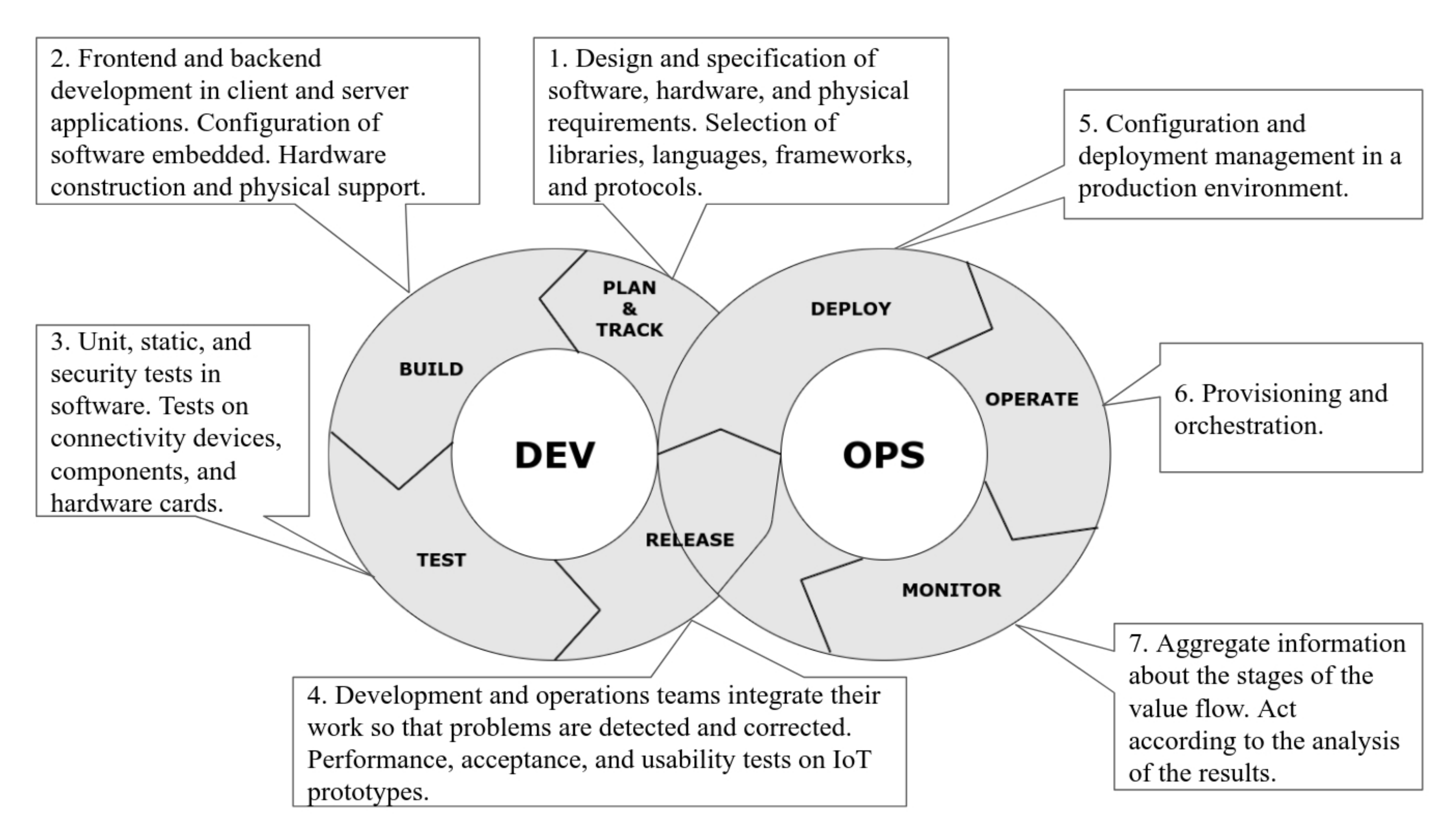}
\caption{Proposed DevOps deployment pipeline for IoT software systems.}~\label{fig:fig6}
\end{figure}

The terms of the pipeline stages can vary between different DevOps implementations. However, the words chosen for this study synthesize each stage according to the discussions of the selected papers, concepts, and tools covered. Figure~\ref{fig:fig6} illustrates a conception DevOps deployment pipeline for IoT software systems according to our analysis of the studies. This model reflects that the use of DevOps in IoT software systems needs continuous experimentation. These stages are not necessarily fixed. There may be others, but these are the ones that have the most discussion in the literature. Following this pipeline with specific tools and techniques generates learning for those involved in the face of challenges and benefits. The path to apply automation to this pipeline must be continuous and incremental, as repetition and practice are prerequisites for the successful use of DevOps in IoT software systems.

It is normal in a search to find an infinity symbol that shows the stages of DevOps. When doing open coding, we noticed evident stages. For this reason, we decided to propose a pipeline that can help anyone who wants to apply DevOps in IoT software systems. It is just a model, an inspiration, only as others that exist for other contexts. However, ours was identified through categorized practices, techniques, and tools about DevOps and IoT software systems. We believe that it is not the same as the others, nor very different, it just represents an instance of the moment in the literature.

We conclude that some tools to release a version operate via a pipeline as code. The pipelines as code technique emphasizes that the configuration of delivery pipelines that build, test, and deploy applications or infrastructure should be treated as code. Ideally, they should be placed under source control and modularized in reusable components with automated testing and deployment. However, the selected studies do not go into much depth on this subject. Experiments on building a pipeline as code could assess the difficulties of this activity, which is a trend in this area.

The studies presented several instances of what may be a deployment pipeline. In general, there was a focus on operating activities. The proposal shown in Figure~\ref{fig:fig6} presents a portrait of the reports made in the selected studies. However, it is necessary to discuss further the requirements for the choice of tools. For instance, we need to search if they have support for the pipeline as code, support for builds with containers, adequate documentation if they are open source, and an active community to help build the pipeline.

In general, we agree with the study by Travassos et al \cite{travassossbes}. The DevOps movement has significant challenges in the state of practice. These challenges may even be present in implementations of IoT projects. However, we believe that scientific research related to the identification of concepts, practices, methods, benefits, and challenges strengthens the body of knowledge to adopt this culture.

We conceive a DevOps deployment pipeline for IoT software systems. Our pipeline contains seven different stages identified according to the concepts and tools used in the studies. This pipeline can be adopted and adapted incrementally according to the needs of each project. We believe that it clarifies the different activities that each stage can handle and then be automated.

\section{Threats to Validity}\label{sec:threats}
We discuss the threat in term of four categories \cite{wohlin}:
\textbf{Construct Validity:} The electronic data sources used may not return all relevant studies. To minimize this threat, we selected five different databases that aggregate papers from various publishers.

\textbf{Internal Validity:} Possible limitations may be related to the judgment of the information presented, which expresses only the authors' point of view. However, frequent meetings were held between all authors to discuss the relevance of papers that should continue in the remaining phases of our analysis.

\textbf{External Validity:} This threat is related to the representativeness of the selected studies published between 2015 and half 2019 to the main goals of this SLR. Our findings on the relationship between DevOps and IoT systems during the selected study period are accurate to the best of our knowledge.

\textbf{Conclusion Validity:} The summary benefits and challenges presents the authors' point of view and may not give the actual definition published by the papers. To minimize this threat, the authors used open coding and discussed every study to reach an agreement.

\section{Related Works}\label{sec:relatedworks}
In this section, we discuss studies related to this work. Some secondary studies in the context of DevOps \cite{devops1,devops2,devops3} and IoT \cite{iot1,iot2} address these matters individually, focusing only on one topic. Other studies deal with more than one theme \cite{devopsembarcados,devopsmicroservicos,iotsofs}. All of these studies are different from our work because we focus on the explicit relationship between DevOps and IoT systems.

Demissie et al. \cite{devopsembarcados} presents a systematic mapping of the use of agile methods in the development on secure embedded systems. The most accomplished combination was Scrum practices with some Extreme Programming practices. Some challenges were mentioned, such as extended time to receive feedback on hardware development, limited understanding of the customer's environment to build the test environment, and the absence of information on the performance of the systems. Our work identifies some challenges encountered with the use of DevOps in IoT systems. These cover not only the client and server, but also the software essential to hardware in embedded systems.

Munoz et al. \cite{devopsmicroservicos} perform a systematic mapping that characterizes nine architectural patterns found in microservices with their advantages and disadvantages. It also highlights different configurations of migration, orchestration, storage, and deployment of this technology in the clouds. The lack of techniques for the DevOps release step is mentioned. Unlike Munoz study, ours identifies some DevOps concepts and tools used in IoT systems that even include the release step.

Maia et al. \cite{iotsofs} discuss scenarios and approaches in the development of IoT-based systems of systems, as well as some research challenges and opportunities in this context. The main results are the identification of studies that propose different architectures for this type of systems and the search for ways to abstract heterogeneity and interoperability between physical devices. Our work presents and discusses as a benefit, the use of cloud computing in IoT systems. In fact, we observed that clouds provide the achievement of different non-functional requirements between devices.

\section{Conclusion} \label{sec:conclusion}
The frequent use of DevOps and IoT software systems is not widely discussed in the literature. In this paper, we present the results of a SLR to identify the explicit relationship between DevOps and IoT software systems. We found twenty-six studies showing five benefits, nine challenges, fifteen concepts, and forty-four tools distributed in seven stages of DevOps for IoT.

The main contributions of this SLR were: (i) identification and discussion of DevOps concepts that are successfully applied to IoT systems; (ii) two lists of benefits and challenges found in the literature on the relationship between DevOps and IoT (Tables~\ref{tab:tab5} and~\ref{tab:tab6}); (iii) two lists of tools and programming languages reported in the studies; and (iv) a set of stages to DevOps deployment pipeline for IoT software systems.

As future work, we suggest a survey and an interview study with developers to understand the relationship between DevOps and IoT software systems from the perspective of professionals in the area. Follow-up studies may be planned under the influence of triangulation of SLR results, survey and interview studies, human-oriented, and tool-guided experiments.

\section*{Acknowledgment}
This research was partially supported by Brazilian funding agencies: CAPES, CNPq, and FAPEMIG.

\end{document}

%% file: tables/Tools.tex
\begin{table}[!h]
\center
\caption{Software tools distributed in the stages of DevOps}\label{tab:tab4}
\begin{tabular}{lp{5.8cm}}
\toprule
Step & Tools \\
\midrule
Plan and Track & Trello, Slack, Hipchat\\
Build & Github, Gitlab, Bitbucket, Git, Maven, Eclipse\\
Test & Cucumber, Sonar, JUnit\\
Release & Codeship, Jenkins, Travis CI, Bamboo, Circle CI, Gitlab CI\\
Deploy & Vagrant, Puppet, Ansible, Terraform, Chef\\
Operate & Rkt, Docker, Swarm, Cloudify, Kubernetes, Mongo DB, Geroku, AWS IoT Core, Hub IoT Azure, Google Cloud IoT, Linux, Openstack\\
Monitor & Nagios, Grafana, Zabbix, Prometheus, Graylog, Elastic\\
\bottomrule
\end{tabular}
\end{table}

%% file: tables/Benefits.tex
\begin{table*}[!h]
\center
\caption{Benefits categories after open coding}\label{tab:tab5}
\begin{tabular}{lll}
\toprule
Benefits & Categories & Studies\\
\midrule
B01 & DevOps framework proposals for IoT & \cite{s24}  \cite{s17} \cite{s20} \cite{s26} \cite{s23} \cite{s13} \cite{s18} \cite{s15} \cite{s05} \cite{s22}\\

B02 & Use of cloud computing as a back-end & \cite{s25} \cite{s10} \cite{s26}\\

B03 & Interoperability with the use of containers & \cite{s08} \cite{s11} \cite{s14}\\

B04 & Microservices induce the formation of small teams & \cite{s12} \cite{s08}\\

B05 & Support platform for solutions that use cellular network & \cite{s16} \cite{s02}\\
\bottomrule
\end{tabular}
\end{table*}

%% file: tables/Challenges.tex
\begin{table*}[!h]
\center
\caption{Challenges categories after open coding}\label{tab:tab6}
\begin{tabular}{lll}
\toprule
Challenges & Categories & Studies\\
\midrule
C01 & Metrics of technological values &  \cite{s02} \cite{s16} \cite{s20} \cite{s06} \cite{s03} \cite{s01}\\

C02 & Heterogeneous and remote device management &  \cite{s09} \cite{s02} \cite{s13} \cite{s04}\\

C03 & Continuous feedback between ops and devs & \cite{s17} \cite{s18}  \cite{s11} \cite{s05}\\

C04 & Address different non-functional requirements & \cite{s12} \cite{s19} \cite{s14} \cite{s22}\\

C05 & Apply a deployment pipeline &  \cite{s11} \cite{s01} \cite{s22}\\

C06 & Continuous documentation & \cite{s20} \cite{s07} \cite{s21}\\

C07 & Need for e-shaped professionals & \cite{s23} \cite{s08} \cite{s21}\\

C08 & Evaluate quality of service & \cite{s02} \cite{s16}\\

C09 & Lack of reports and devops frameworks in IoT & \cite{s10} \cite{s23}\\
\bottomrule
\end{tabular}
\end{table*}